\documentclass[letterpaper]{article} 
\usepackage{booktabs}
\usepackage{multirow}
\usepackage{aaai2026}  
\usepackage{times}  
\usepackage{helvet}  
\usepackage{courier}  
\usepackage[hyphens]{url}  
\usepackage{graphicx} 
\usepackage{natbib}  
\usepackage{caption} 
\newcommand{\algoname}{PanFoMa}
\usepackage{algorithm}
\usepackage{algorithmic}
\usepackage{amsmath}
\usepackage{amssymb}
\usepackage{newfloat}
\usepackage{listings}
\DeclareCaptionStyle{ruled}{labelfont=normalfont,labelsep=colon,strut=off} 
\floatstyle{ruled}
\newfloat{listing}{tb}{lst}{}
\floatname{listing}{Listing}
\title{PanFoMa: A Lightweight Foundation Model and Benchmark for Pan-Cancer}
\author {
    Xiaoshui Huang\textsuperscript{\rm 1}\equalcontrib,
    Tianlin Zhu\textsuperscript{\rm 2}\equalcontrib,
    Yifan Zuo\textsuperscript{\rm 2}\thanks{Corresponding author.},
    Xue Xia\textsuperscript{\rm 2},
    Zonghan Wu\textsuperscript{\rm 4},
    Jiebin Yan\textsuperscript{\rm 2},
    Dingli Hua\textsuperscript{\rm 2},
    Zongyi Xu\textsuperscript{\rm 5},
    Yuming Fang\textsuperscript{\rm 2},
    Jian Zhang\textsuperscript{\rm 3}
}
\affiliations {
    \textsuperscript{\rm 1}Shanghai Jiao Tong University\\
    \textsuperscript{\rm 2}Jiangxi University of Finance and Economics\\
    \textsuperscript{\rm 3}University of Technology Sydney\\
    \textsuperscript{\rm 4}East China Normal University\\
    \textsuperscript{\rm 5}Chongqing university of posts and telecommunications\\
}

\usepackage{bibentry}

\begin{document}

\maketitle

\begin{abstract}
Single-cell RNA sequencing (scRNA-seq) is essential for decoding tumor heterogeneity. However, pan-cancer research still faces two key challenges: learning discriminative and efficient single-cell representations, and establishing a comprehensive evaluation benchmark. In this paper, we introduce \algoname, a lightweight hybrid neural network that combines the strengths of Transformers and state-space models to achieve a balance between performance and efficiency. \algoname~consists of a front-end local-context encoder with shared self-attention layers to capture complex, order-independent gene interactions; and a back-end global sequential feature decoder that efficiently integrates global context using a linear-time state-space model. This modular design preserves the expressive power of Transformers while leveraging the scalability of Mamba to enable transcriptome modeling, effectively capturing both local and global regulatory signals. To enable robust evaluation, we also construct a large-scale pan-cancer single-cell benchmark, \algoname Bench, containing over 3.5 million high-quality cells across 33 cancer subtypes, curated through a rigorous preprocessing pipeline. Experimental results show that \algoname~outperforms state-of-the-art models on our pan-cancer benchmark (+4.0\%) and across multiple public tasks, including cell type annotation (+7.4\%), batch integration (+4.0\%) and multi-omics integration (+3.1\%). The code is available at \texttt{https://github.com/Xiaoshui-Huang/PanFoMa}.
\end{abstract}


\section{Introduction}
The revolutionary advances in single-cell RNA sequencing (scRNA-seq) technology have provided an unprecedentedly powerful tool for systematically dissecting the heterogeneity of complex biological systems, such as tumors, at single-cell resolution \cite{jovic2022single}. By precisely capturing the gene expression profile of each cell, we can gain deep insights into the underlying mechanisms of tumor initiation, progression, metastasis, and response to therapy. Therefore, developing computational models capable of learning effective representations of cells and genes from high-dimensional, sparse transcriptomic data has become a central challenge in computational biology. The deep representations learned by such models are foundational not only for advancing precision medicine and personalized diagnostics \cite{dutta2022single}, but also for a broad range of applications, including biomarker discovery, drug target identification, and fundamental studies of cellular processes \cite{van2023applications}.
\begin{figure}
    \centering
    \includegraphics[width=0.9\linewidth]{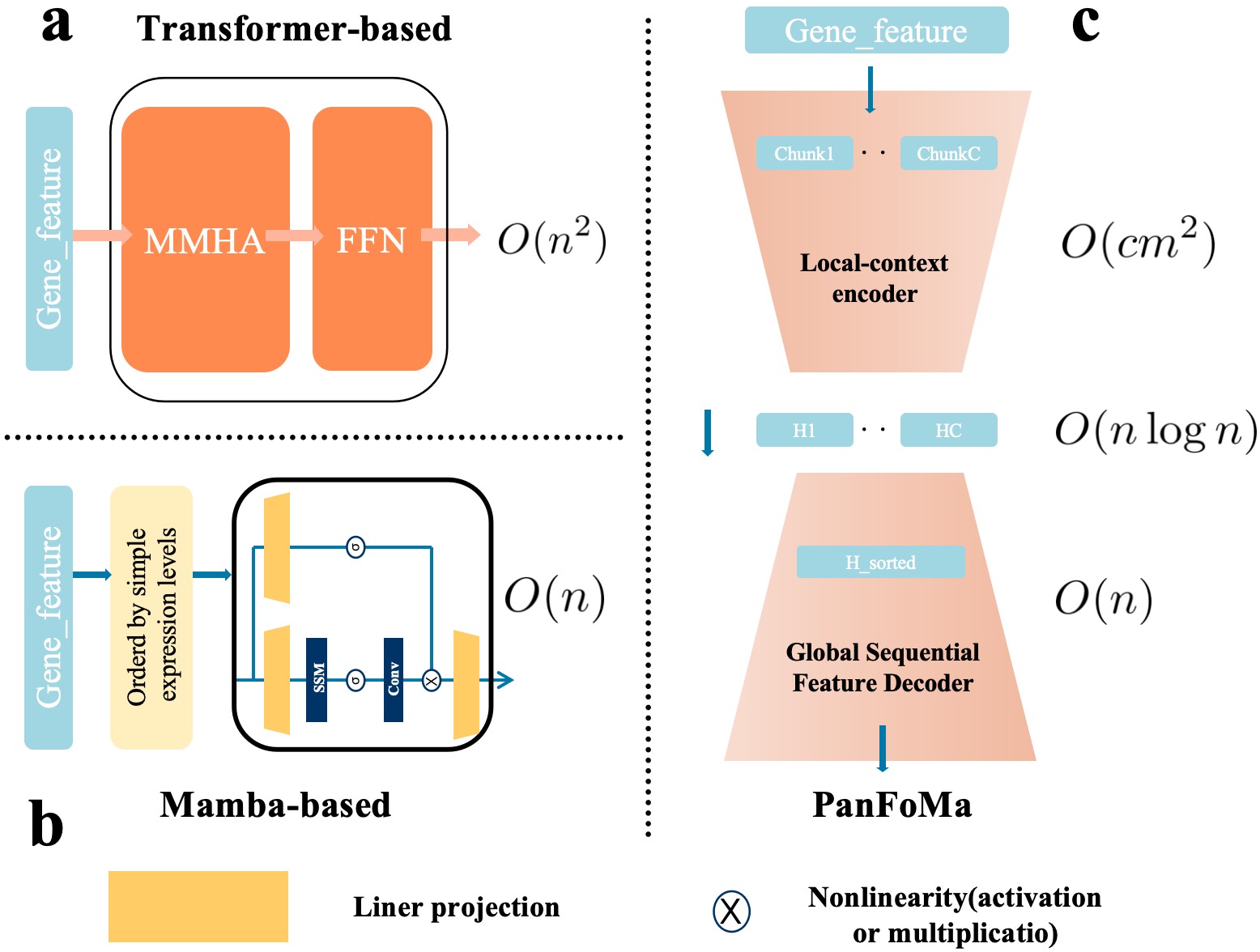}
    \caption{Comparison of different architectures for modeling single-cell gene expression. (a) Transformer-based methods capture local features but incur $O(n^2)$ complexity. (b) Mamba-based models offer $O(n)$ efficiency but require fixed input ordering. (c) Our proposed PanFoMa captures both local and global dependencies via a hybrid encoder-decoder design, resulting in an overall computational complexity of $O(C \cdot M^2 + N \log N)$, where $N=C \cdot M$.} 
    \label{fig:1}
\end{figure}

Most existing single-cell foundation models are transformer-based \cite{he2024integrated, cui2024scgpt, hao2024large, theus2024cancerfoundation, cui2024geneformer, adduri2025predicting, fang2025cell, zeng2025cellfm}, with scGPT \cite{cui2024scgpt} being a representative example. Inspired by advances in natural language processing, these models conceptualize genes as "tokens" and cells as "sentences," leveraging the self-attention mechanism to capture complex dependencies among genes. Through generative pretraining on massive unlabeled single-cell datasets, they aim to learn rich and nuanced representations of the "gene language."
However, these models face several inherent limitations. First, the \textit{computational complexity} of the self-attention mechanism scales quadratically with the number of input genes, making it computationally prohibitive to process the complex transcriptome, which can contain tens of thousands of genes. Consequently, current approaches \cite{cui2024scgpt, cui2024geneformer} typically process only a subset of genes (e.g., 2048), selected via top-K highly variable genes (HVGs) \cite{cui2024scgpt, zeng2025cellfm}, to reduce computational costs—at the expense of capturing only localized gene interactions.
Second, this HVGs-based gene selection strategy \cite{cui2024scgpt, zeng2025cellfm} has notable drawbacks: it may \textit{exclude important low-expression functional genes} (e.g., transcription factors) and introduces \textit{analytical bias} into downstream tasks. These limitations compromise the models’ ability to generalize across diverse biological contexts and hinder their interpretability in functional genomics applications.

To overcome the limitations of above tranformer-based methods, recently, researchers have begun exploring the state-space models \cite{qi2025bidirectional} for learning single-cell representations. Mamba \cite{gu2023mamba} offers linear-time complexity by maintaining a latent state that sequentially aggregates input information, thus potentially capturing global dependencies more efficiently than Transformers. However, directly applying Mamba to single-cell data presents several critical challenges. First, Mamba  is fundamentally designed for ordered sequences, whereas a cell's gene expression profile is naturally an unordered set—gene interactions do not depend on any inherent sequence. To apply Mamba, existing approaches must impose a heuristic and fixed ordering of genes (e.g., by mean expression), thereby forcing an artificial sequence structure. \textit{This static ordering is suboptimal}, as it neglects the context-dependent nature of gene functions, which emerge from complex and dynamic regulatory networks rather than simple expression levels. As a result, such ordering may impair the model's ability to capture authentic biological relationships. Second, although Mamba’s latent state enables information integration across input sequences, its fixed dimensionality suffers from forgetting distant input tokens, which can \textit{limit the ability to retain long-range dependencies}, particularly when dealing with long gene sequences. This constraint poses a serious challenge in tasks like Pan-Cancer classification and analysis, where capturing broad, global patterns across the transcriptome is essential. The limited memory capacity may result in degraded performance when modeling distant gene interactions critical for accurate diagnosis and interpretation.

To overcome the above limitations and achieve both high accuracy and computational efficiency in Pan-Cancer analysis, we propose a novel computational paradigm based on a decoupled modeling strategy for single-cell transcriptomes. Specifically, we separate the tasks of capturing local gene interactions and integrating global information, and assign them to dedicated architectures. A lightweight Transformer encoder captures local, context-rich representations from a limited gene subset, while a Mamba-based decoder integrates these into a global cell-state representation using its efficient latent-state mechanism.
This division of labor combines the expressive power of Transformers with the linear scalability of Mamba, resolving the trade-off between performance and efficiency. Biologically, it mimics the information flow from complex local regulatory signals to a unified cell state, enabling end-to-end modeling of the transcriptome.

We implement this idea in a innovative hybrid neural network named \algoname, which consists of two core modules. First, the \textbf{Local-context Encoder} partitions the input of thousands of genes into multiple fixed-size chunks and processes these chunks in parallel using a lightweight Transformer with shared parameters across layers. This design efficiently captures the complex interaction relationships within each chunk while significantly reducing computational and memory overhead. Second, the \textbf{Global Sequential Feature Decoder} introduces a novel processing pipeline: it first synthesizes a global vector representing the overall cell state by average pooling the summary information (\texttt{cls\_token}) from all chunks. Next, it leverages this global vector to dynamically compute an "importance" score for each gene via dot product and intelligently sorts all genes accordingly. Finally, this biologically meaningful, ordered sequence is fed into a bidirectional Mamba module for deep integration, and the output is produced through a gated mechanism.

In addition, existing pan-cancer datasets \cite{weinstein2013cancer,vstancl2023machine,nofech2023pan} often suffer from limited subtype coverage and sample size, introducing bias and overfitting in model evaluation. To address this, we construct and release a large-scale, high-quality benchmark dataset comprising over 3.5 million single cells across 33 major cancer types, following a rigorous data curation pipeline. This challenging benchmark requires strong generalization and is designed to drive progress in pan-cancer foundation model development.

In summary, the contributions are three aspects:
\begin{itemize}
    \item We design a hierarchical \textit{hybrid architecture} that integrates the strengths of Transformer-based local modeling and Mamba-based global integration, achieving a superior balance between performance and computational scalability. 
    \item We develop a \textit{global-informed dynamic sorting mechanism} that adaptively determines gene input sequences based on each cell's global transcriptomic context, moving beyond fixed heuristic rules and enhancing biological relevance.
    \item We construct a comprehensive \textit{pan-cancer benchmark}, and through extensive evaluations on both curated and public benchmarks, we demonstrate that our model consistently outperforms existing state-of-the-art approaches, providing a robust framework to accelerate advances in precision medicine.
\end{itemize}

\section{Related work}
\subsection{Single cell foundation models}
In recent years, drawing inspiration from successful paradigms in Natural Language Processing (NLP), Foundation Models have been widely applied to single-cell transcriptomics, aiming to learn \textit{universal} cell and gene representations through large-scale self-supervised pre-training. The development in this field has primarily evolved along two core architectural lines: early models based on Transformers \cite{cui2024geneformer,cui2024scgpt,zeng2025cellfm,fang2025cell,huang2024frozen, huang2025psreg}, and a new generation of architectures based on State Space Models (SSMs) \cite{qi2025bidirectional} that have emerged in pursuit of higher efficiency.

The main idea of transformer-based methods leverages its powerful self-attention mechanism to capture complex, long-range interactions between genes. scGPT \cite{cui2024scgpt} is one of the representative works in this area. It adopts a GPT-like generative pre-training framework \cite{radford2018improving,zheng2024point,mei2022unsupervised}, designed to complete the entire gene expression profile based on partial gene or cell prompts. To address the fundamental challenge of the unordered nature of gene expression data, scGPT designed a special attention mask mechanism, which simulates an iterative generation process by randomly partitioning known and unknown genes during training, thereby enabling autoregressive-style learning on non-sequential data. 
In contrast, GeneFormer \cite{cui2024geneformer} introduces a rank-based input encoding and masked gene prediction strategy to mitigate batch effects and better capture gene regulatory networks. He et al. \cite{he2024integrated} further fine-tuned GeneFormer for cancer applications by extending its pretraining with approximately 14 million cancer-derived cells. scFoundation \cite{hao2024large} adopts an asymmetric transformer-like architecture to model complex gene relationships across diverse cell types. CancerFoundation \cite{theus2024cancerfoundation} focuses exclusively on malignant cells during training. Most recently, CellFM \cite{fang2025cell} presents a single-cell foundation model with 800 million parameters, trained on 100 million human cells using a modified RetNet framework to balance efficiency and performance.

Despite the success of above Transformer-based models, they share an unavoidable bottleneck: the quadratic computational complexity,  excluding low-expression function genes and analytical bias. To break through the first bottleneck, Mamba \cite{gu2023mamba} has provided a new solution for efficiently processing sequence data with linear computational complexity by using a latent state to accumulate the previous input tokens and propagate to the next token prediction. GeneMamba \cite{qi2025bidirectional} is a Mamba-based foundation model specifically designed for single-cell transcriptomics. It inherits the expression-based ranking input strategy from GeneFormer but replaces the core computational backbone from a Transformer to a bidirectional Mamba (Bi-Mamba). Although Mamba-based architectures achieve a breakthrough in computational efficiency, their performance is highly dependent on a fixed, heuristic rule for creating a serialized input. However, this static sorting mechanism is suboptimal because it overlooks a core biological fact: a gene's importance is not static but is determined by its dynamic functional role within a specific cellular context. Therefore, the field currently lacks a model that can simultaneously achieve a deep understanding of unordered gene relationships and highly efficient serialized processing.

\subsection{Pan-cancer prediction}

 Pan-cancer prediction is a key area in cancer research, aiming to leverage machine learning and deep learning to identify shared and distinct features across different cancer types for improved diagnosis , prognosis, and treatment \cite{nopour2024prediction,yang2025comprehensive}. Zhang et al. \cite{zhang2025single} proposed a unified multimodal framework for pan-cancer survival prediction by integrating whole-slide histopathology images and gene expression data, addressing the generalization gap across cancer types. Bjerregaard-Michelsen et al. \cite{bjerregaard2025machine}  compared pan-cancer and single-cancer models for short-term mortality prediction, highlighting the utility of ML in treatment decisions but with limited long-term forecasting power. Ocasio et al. \cite{ocasio2024pan} developed a transcriptome-based deep learning model for pan-cancer classification using TCGA data, though limited by dataset bias. Lareau et al. \cite{lareau2021charting}  showed that integrating multiple datasets improves drug sensitivity prediction, while challenges remain in modeling drug resistance. Salmanpour et al. \cite{Salmanpour_2025} evaluates 226 studies applying machine learning to PET and SPECT imaging for cancer outcome prediction, revealing that deep radiomics and fusion models yield the best performance, with PET-based approaches outperforming SPECT. Ouyang et al. \cite{ouyang2025global} reviewed existing models for drug response prediction, pointing out current limitations and the need for more practical guidance on model development. However, the current existing methodds face challenges in multimodal data integration and model generalization \cite{singh2025comprehensive}. At the same time, it is crucial to address data bias in the training dataset and benchmarks by using diverse datasets for training and validation, thereby enhancing the generalizability of the models. This paper focus on these challenges by proposing a foundation model and a novel pan-cancer benchmark.

\section{Method}
\subsection{Overview}
\begin{figure*}[ht]
    \centering
    \includegraphics[width=\linewidth]{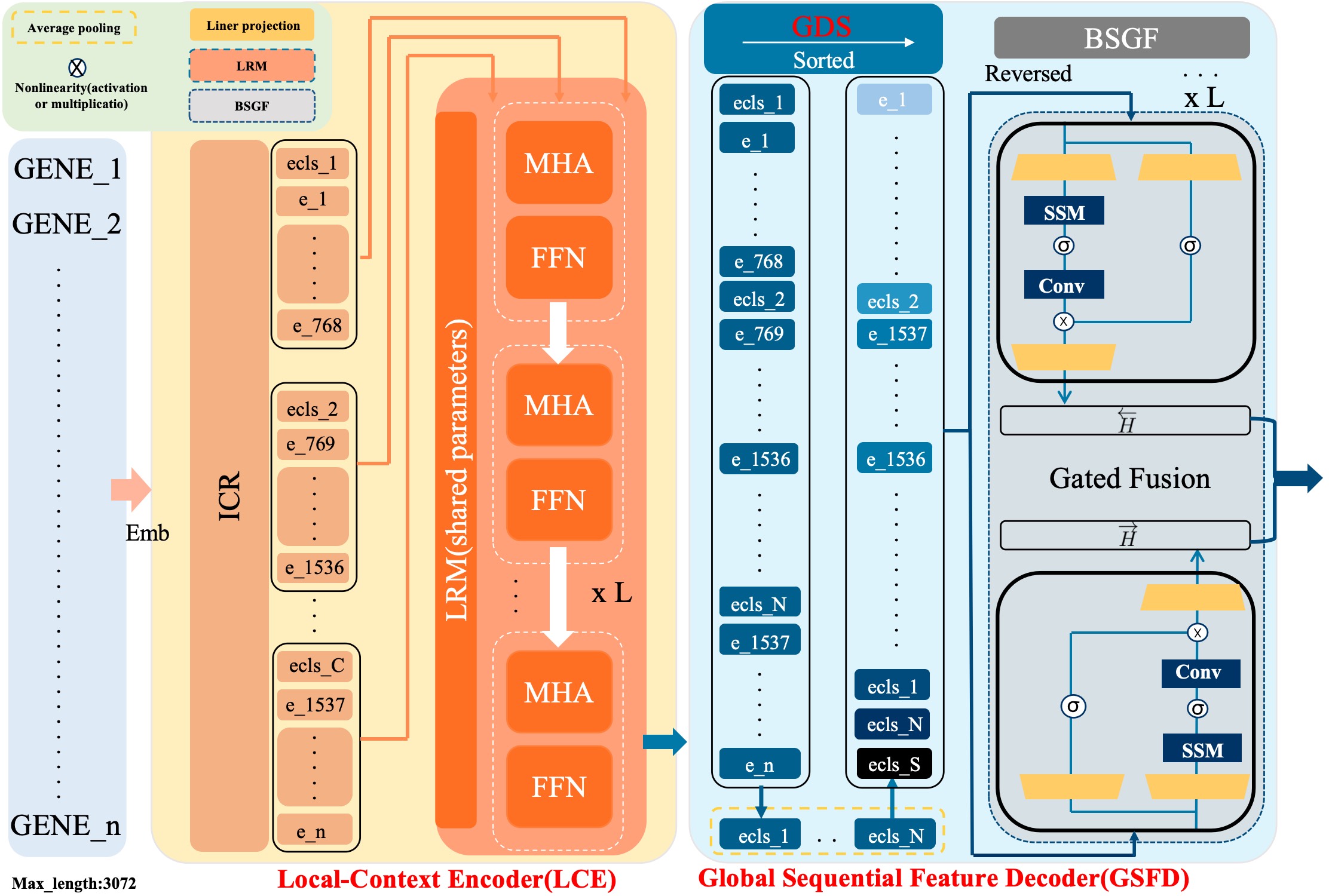}
    \caption{Overview of the \algoname~ algorithm. Firstly, the \textbf{Local-context encoder} first divide the input into chunks and encodes local gene interactions within transcriptome chunks using shared Transformer layers. Second, the \textbf{global sequential feature decoder} guides dynamic reordering of all genes based on relevance. The reordered sequence is processed by a bidirectional Mamba decoder, and a gating module fuses features to generate the final cell representation.  }
    \label{fig:overveiw}
\end{figure*}

To address the dual challenges of accurate and efficient in modeling whole-transcriptome single-cell data, we propose \algoname, a hybrid neural network based on a hierarchical local-to-global processing paradigm. The core idea is to decompose the transcriptome into manageable local blocks for parallel contextual encoding, followed by global integration via dynamic gene reordering.
By unifying localized deep modeling and globally informed integration, \algoname~ provides a scalable and effective solution for high-resolution single-cell representation learning. The overall framework is illustrated in Figure~\ref{fig:overveiw}.

\subsection*{Module 1: Local-context encoder}

This module is designed to efficiently process large-scale gene inputs in parallel and to learn a feature representation for each gene that is rich in information about its local regulatory environment. Its design follows a "\textbf{divide and conquer}" strategy, decomposing a large-scale global relationship modeling problem into multiple manageable local problems. It consists of the following two key steps:

\subsubsection*{Input chunking and representation (ICR)}

To overcome the quadratic ($O(N^2)$) computational and memory complexity bottleneck faced by standard Transformers when processing ultra-long gene sequences, we first partition the input into chunks. In each training epoch, we randomly sample 3072 genes from a cell and divide them into $C=4$ non-overlapping chunks, with each chunk $k$ containing a fixed size of $M=768$ genes. To independently learn and aggregate summary information for each chunk, we prepend an independent, learnable token, \texttt{[CLS]}, to the beginning of each chunk.

For each gene $g_{k,i}$ in chunk $k$, we project its discrete gene ID and binned expression value through separate embedding layers and then fuse them via element-wise addition to obtain its input feature embedding $e_{k,i} \in \mathbb{R}^D$:
\begin{equation}
    e_{k,i} = \text{Emb}_{\text{id}}(g_{\text{id}_{k,i}}) + \text{Emb}_{\text{val}}(g_{\text{val}_{k,i}})
\end{equation}
where $D$ is the embedding dimension. Thus, the complete input sequence for the $k$-th chunk is a matrix $S_k \in \mathbb{R}^{(M+1) \times D}$, represented as $S_k = \{e_{\texttt{[CLS]}_k}, e_{k,1}, \dots, e_{k,M}\}$.

\subsubsection*{Local relationship modeling (LRM)}

All $C$ chunks are then processed in parallel by a lightweight Transformer encoder, which consists of a six-layer stack ($L=6$) with shared parameters across layers. Let the input to layer $l$ for chunk $k$ be $H_k^{(l-1)} \in \mathbb{R}^{(M+1) \times D}$ (where $H_k^{(0)} = S_k$). The output of the layer, $H_k^{(l)}$, is computed by applying the standard Transformer block operations:
\begin{equation}
    H_k^{(l)} = \text{TransformerBlock}(H_k^{(l-1)})
\end{equation}
After six layers, the module outputs a final hidden state matrix $H_k^{(L)}$ for each chunk. The output of the local-context encoder is twofold: first, the gene-specific embeddings $H_{\text{genes},k}^{(L)} \in \mathbb{R}^{M \times D}$, which are the rows of $H_k^{(L)}$ corresponding to the gene tokens. Second, the \texttt{cls\_token} embedding $h_{\texttt{[CLS]},k}^{(L)} \in \mathbb{R}^D$, which is the first row of $H_k^{(L)}$ and serves as a summary vector for the chunk.

\subsection*{Module 2: Global sequential feature decoder}

The core task of this module is to perform global information integration, dynamic sorting, and deep sequential processing based on the locally-informed features generated by the local-context encoder.

\subsubsection*{Global-informed dynamic sorting (GDS)}

First, we synthesize a global cell state vector, $h_{\text{global\_cls}} \in \mathbb{R}^D$, by applying \textbf{Average Pooling} to the \texttt{cls\_token} summary vectors from all $C$ chunks:
\begin{equation}
    h_{\text{global\_cls}} = \frac{1}{C} \sum_{k=1}^{C} h_{\texttt{[CLS]},k}^{(L)}
\end{equation}
Next, we concatenate the gene-specific embeddings from all chunks to form a single matrix $H_{\text{genes,full}}^{(L)} \in \mathbb{R}^{N \times D}$, where $N = C \times M$. We then compute an importance score $s_i$ for each gene embedding $h_i \in H_{\text{genes,full}}^{(L)}$ via its dot product with the global cell state vector:
\begin{equation}
    s_i = h_i \cdot h_{\text{global\_cls}}^T
\end{equation}
Based on the score vector $s = \{s_1, \dots, s_{N}\}$, we sort the rows of $H_{\text{genes,full}}^{(L)}$ in descending order to obtain the sorted feature matrix $H_{\text{sorted}}^{(L)}$.

\subsubsection*{Bidirectional scanning and gated fusion (BSGF)}

The sorted gene sequence, $H_{\text{sorted}}^{(L)}$, is then fed into a six-layer bidirectional Mamba module. Each layer consists of a forward and a backward Mamba:
\begin{align*}
    \overrightarrow{H}_{\text{mamba}} &= \text{Mamba}_{\text{fwd}}(H_{\text{sorted}}^{(L)}) \\
    \overleftarrow{H}_{\text{mamba}} &= \text{Mamba}_{\text{bwd}}(\text{Reverse}(H_{\text{sorted}}^{(L)}))
\end{align*}
where $\overrightarrow{H}_{\text{mamba}}, \overleftarrow{H}_{\text{mamba}} \in \mathbb{R}^{N \times D}$ are the output hidden state sequences.

To fuse these bidirectional flows, a \textbf{gated fusion mechanism} learns a gate vector $\gamma_i \in \mathbb{R}^D$ for each gene's sorted feature $h_{\text{sorted},i} \in H_{\text{sorted}}^{(L)}$:
\begin{equation}
    \gamma_i = \sigma(\text{Linear}(h_{\text{sorted},i}))
\end{equation}
where $\sigma$ is the Sigmoid function. This gate is used to compute the final fused feature $h_{\text{fused}, i}$ as a weighted sum of the forward and backward hidden states, $\overrightarrow{h}_{\text{mamba},i}$ and $\overleftarrow{h}_{\text{mamba},i}$:
\begin{equation}
    h_{\text{fused}, i} = \gamma_i \odot \overrightarrow{h}_{\text{mamba},i} + (1 - \gamma_i) \odot \overleftarrow{h}_{\text{mamba},i}
\end{equation}
where $\odot$ denotes element-wise multiplication. This approach allows the model to adaptively weigh the contextual information from both directions for each gene.

\begin{figure*}[h]
    \centering
    \includegraphics[width=0.9\linewidth]{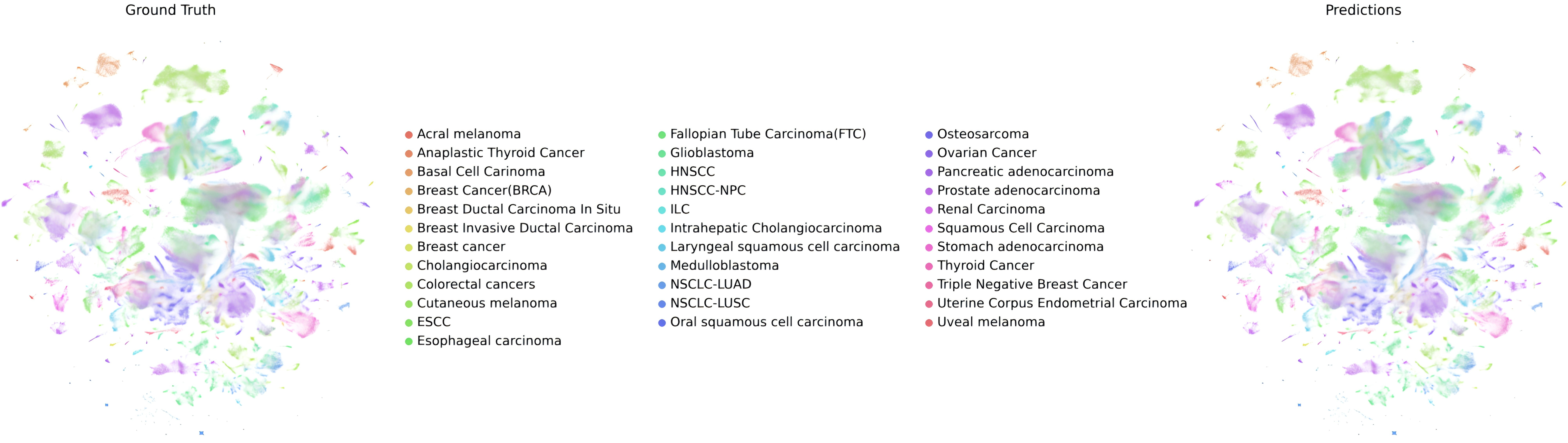}
    \caption{Visualization of pan-cancer classification Results. The \algoname~can clearly separate the different cancer subtypes. }
    \label{fig:cancls}
\end{figure*}
\subsection{Pan-Cancer benchmark construction}
To construct a diverse and high-quality Pan-Cancer benchmark, named as \textit{\algoname Bench}, we followed a rigorous and systematic workflow. We began by extensively searching and curating publicly available human cancer single-cell transcriptomic datasets from the National Center for Biotechnology Information (NCBI) database. Selected datasets were required to include essential metadata, such as cellular transcriptomic profiles, cancer type, tissue of origin, and corresponding patient information. To ensure consistency and enable cross-dataset integration, all gene identifiers were standardized to official symbols defined by the HUGO Gene Nomenclature Committee (HGNC).

After standardization, the individual datasets were merged to form a unified Pan-Cancer cohort. This integrated dataset was then subjected to a \textit{comprehensive quality control (QC) pipeline to ensure data fidelity and reliability}. The QC process involved the following filtering steps:
\begin{itemize}
\item (1) Removal of cells with a low number of expressed genes;
\item (2) Exclusion of potential doublets or multiplets, identified by abnormally high gene or UMI counts;
\item (3) Filtering out cells with a high proportion of mitochondrial gene expression, indicative of low viability;
\item (4) Elimination of genes with low expression across the dataset to reduce noise and improve computational efficiency.
\end{itemize}

To support the training and validation of our model, and to contribute a valuable resource to the pan-cancer research community, we constructed a large-scale, high-quality pan-cancer single-cell transcriptomic dataset. We began by conducting a systematic literature search in public databases, including the National Center for Biotechnology Information (NCBI), using keywords such as "pan-cancer" and "cancer". This search process ultimately identified 83 relevant published studies, from which the publicly available single-cell sequencing data were sourced for our dataset.Following rigorous data filtering, standardized preprocessing pipelines, and uniform quality control, we successfully integrated and constructed this high-quality benchmark dataset, which was used for all subsequent analyses. In total, our curated dataset comprises approximately 3.5 million high-quality cells from 616 patients, comprehensively covering 33 distinct cancer subtypes and originating from 23 different tissue types. 


\section{Experiments}
\subsection{Experiment setting}
This section aims to comprehensively estimate the proposed \algoname~ from two aspects: Pan-cancer diagnosis and the ability of foundation models. The Pan-cancer diagnosis is evuated by the task of Pan-cancer classification, Gene regularoty network inference; and the ability of foundation model is evaluated by the tasks of Batch integration, cell type annotation and multi-omic integration.

We compare our methods with the recent proposed single-cell foundation models to demonstrate our better cell representation learning ability: GeneFormer \cite{cui2024geneformer}, scGPT \cite{cui2024scgpt}, scFoundation \cite{hao2024large} and GeneMamba \cite{qi2025bidirectional}.

\subsection{Pan-cancer diagnosis}

Pan-Cancer diagnosis is a cancer subtype classification problem \cite{ocasio2024pan}. To assess the performance of our proposed \algoname~ on the Pan-Cancer diagnosis task, we fine-tuned the model on the training datasets of our benchmark and evaluated its classification accuracy on the corresponding test sets. For comparison, four state-of-the-art single-cell foundation models were evaluated under identical settings.

\begin{table}[ht]
\centering

\begin{tabular}{@{}llcc@{}}
\toprule
\textbf{Datasets} & \textbf{Models} & \textbf{ACC} & \textbf{Macro-F1} \\
\midrule
\multirow{5}{*}{PanFoMaBench} & GeneFormer   & 0.9124 & 0.8851 \\
                            & scGPT        & 0.9013 & 0.8732 \\
                            & scFoundation & 0.8876 & 0.8491 \\
                            & GeneMamba    & 0.9026 & 0.8619 \\
                            & \textbf{Ours}         & \textbf{0.9474} & \textbf{0.9250} \\
\bottomrule
\end{tabular}
\caption{Pan-Cancer diagnosis on our benchmark.}
\label{tab:pancancer_results}
\end{table}

As shown in Table~\ref{tab:pancancer_results}, \algoname~ achieved an accuracy of 94.74\%, outperforming the second-best model, GeneFormer, by 3.5\%. In terms of macro-F1 score, \algoname~ attained 92.5\%, representing a 4\% improvement over GeneFormer. Furthermore, our model consistently outperformed GeneMamba in both accuracy and F1 score. These findings underscore the importance of integrating both global and local contextual information for accurate Pan-Cancer diagnosis. The visual results show in Figure \ref{fig:cancls}, which show that the proposed \algoname~ can clearly separate the different cancer subtypes and extremely similar to Groundtruth.

\subsection{Gene regulatory network inference}

The gene regulatory network (GRN) inference task \cite{cui2024scgpt} aims to reverse-engineer regulatory relationships from single-cell transcriptomic data, thereby revealing key regulatory circuits that govern specific cell states or disease processes. An effective computational model should accurately capture strong gene–gene interactions and construct networks that align with established biological knowledge.

\begin{figure}[h]
\centering
\includegraphics[width=0.9\linewidth]{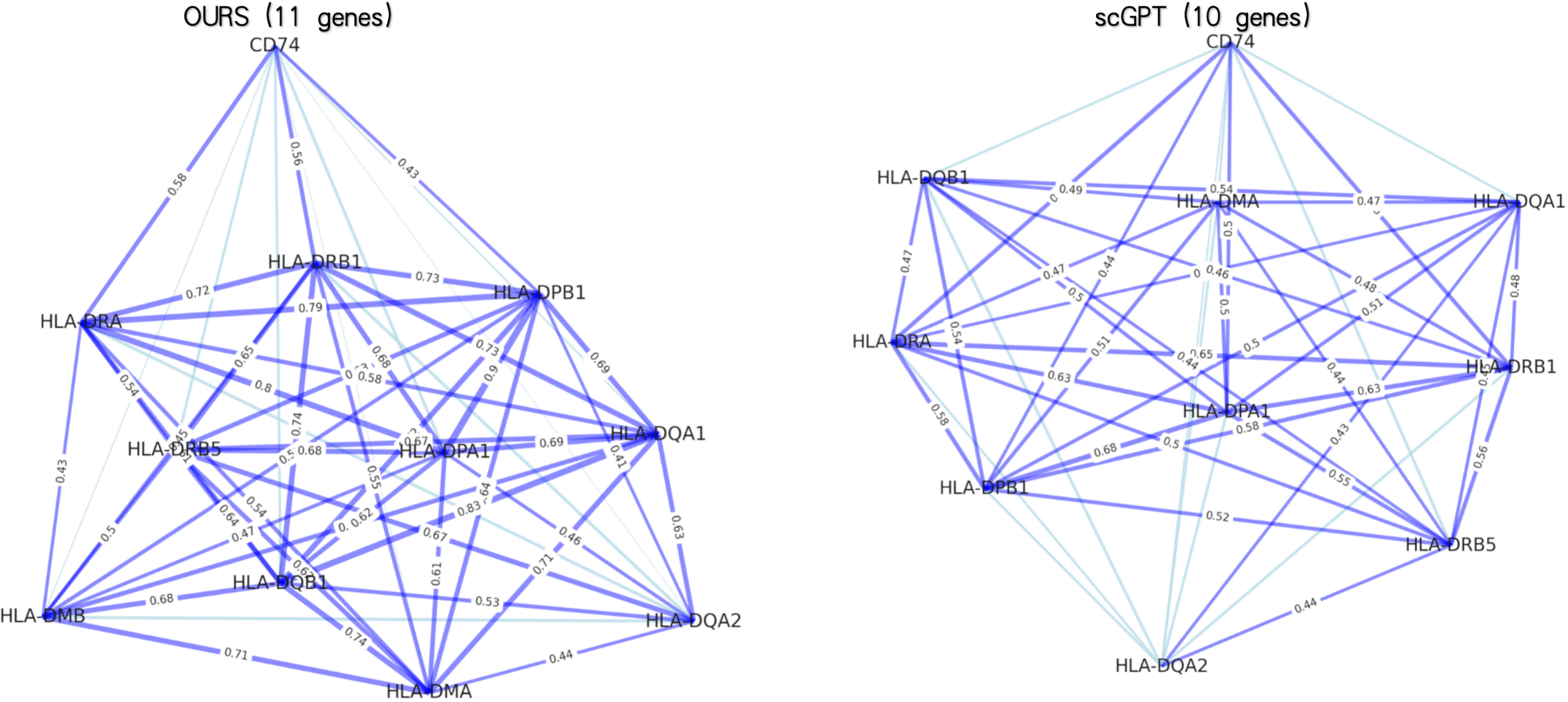}
\caption{Visual comparison of GRN.}
\label{fig:grn}
\end{figure}

To intuitively assess the capability of \algoname~ in capturing gene–gene regulatory interactions, we utilize its attention mechanism to infer GRNs. Specifically, we focus on genes associated with major histocompatibility complex (MHC) class II molecules, which play a central role in immune cell function. As shown in Figure~\ref{fig:grn}, we compare the network inferred by \algoname~ with that of the baseline model, scGPT. Our model not only identifies gene interactions with higher confidence but also recovers one additional relevant gene compared to scGPT.

\subsection{Batch integration}

To evaluate batch integration performance, we follow the protocol in \cite{qi2025bidirectional}, adopting two widely used metrics: $Avg\_batch$ (assessing batch effect removal) and $Avg\_bio$ (measuring biological variance preservation). Higher scores indicate better performance. We benchmark \algoname~ against four advanced models on five challenging, multi-batch public datasets. As summarized in Table~\ref{tab:benchmark_results}, \algoname~ consistently outperforms both transformer-based and Mamba-based baselines in most cases, demonstrating its superior capability in integrating multi-batch single-cell data.
\begin{table}[ht]
\centering

\resizebox{\linewidth}{!}{%
\begin{tabular}{@{}llcccccc@{}}
\toprule
\textbf{Metric} & \textbf{Dataset} & \textbf{GeneFormer} & \textbf{scGPT} & \textbf{scFoundation} & \textbf{GeneMamba} & \textbf{Ours} \\
\midrule
{Avg\_batch} & Immune           & 0.8153 & 0.9194 & 0.8904 & 0.9536 &\textbf{ 0.9641} \\
                            & PBMC12k          & 0.9545 & \textbf{0.9755} & 0.9628 & 0.9604 & 0.9701 \\
                            & BMMC             & 0.7720 & 0.8431 & 0.7598 & 0.9157 & \textbf{0.9312 }\\
                            & Perirhinal Cortex& 0.9127 & 0.9600 & 0.9560 & \textbf{0.9673} & 0.9661 \\
                            & Covid-19         & 0.8240 & 0.8625 & 0.8346 & 0.8742 & \textbf{0.9173} \\
\midrule
{Avg\_bio}   & Immune           & 0.6983 & 0.7879 & 0.7337 & 0.8131 & \textbf{0.8332} \\
                            & PBMC12k          & 0.7891 & \textbf{0.9018} & 0.8662 & 0.8344 & 0.8631 \\
                            & BMMC             & 0.6324 & 0.6576 & 0.5250 & 0.7628 & \textbf{0.8021} \\
                            & Perirhinal Cortex& 0.8547 & 0.9552 & \textbf{0.9606} & 0.9062 & 0.9598 \\
                            & Covid-19         & 0.5567 & \textbf{0.6476} & 0.5468 & 0.5537 & 0.6338 \\
\bottomrule
\end{tabular}%
}
\caption{Comparison of batch integration across datasets.}
\label{tab:benchmark_results}
\end{table}

\subsection{Cell type annotation}

Cell type annotation \cite{chen2023transformer} aims to assign a known biological identity label to each cell based on its unique gene expression profile. To comprehensively evaluate the performance of our proposed \algoname~ on the cell type annotation task, we selected four widely used public benchmark datasets \cite{qi2025bidirectional, hoadley2018cell}—hPancreas, MS, Myeloid, and Myeloid\_b—and adopted Accuracy and Macro-F1 score as the primary evaluation metrics.

\begin{table}[h]

\begin{tabular}{@{}llcc@{}}
\toprule
\textbf{Datasets} & \textbf{Models} & \textbf{Acc} & \textbf{Macro-F1} \\
\midrule
\multirow{5}{*}{hPancreas} & GeneFormer   & 0.9665 & 0.7450 \\
                          & scGPT        & 0.9710 & 0.7632 \\
                          & scFoundation & 0.9602 & 0.7101 \\
                          & GeneMamba    & 0.9713 & 0.7710 \\
                          & Ours         & \textbf{0.9815} & \textbf{0.7760} \\
\midrule
\multirow{5}{*}{MS}        & GeneFormer   & 0.7650 & 0.6220 \\
                          & scGPT        & 0.8471 & 0.6630 \\
                          & scFoundation & 0.7763 & 0.6812 \\
                          & GeneMamba    & 0.6825 & 0.5342 \\
                          & Ours         & \textbf{0.8563} & \textbf{0.7016} \\
\midrule
\multirow{5}{*}{Myeloid}   & GeneFormer   & 0.6445 & 0.3600 \\
                          & scGPT        & 0.6341 & 0.3562 \\
                          & scFoundation & 0.6446 & 0.3646 \\
                          & GeneMamba    & \textbf{0.6607} & \textbf{0.3650} \\
                          & Ours         & 0.6515 & 0.3529 \\
\midrule
\multirow{5}{*}{Myeloid\_b} & GeneFormer   & 0.9540 & 0.9380 \\
                          & scGPT        & 0.9421 & 0.9434 \\
                          & scFoundation & 0.9574 & 0.9569 \\
                          & GeneMamba    & 0.9603 & 0.9235 \\
                          & Ours         & \textbf{0.9726} & \textbf{0.9351} \\
\bottomrule
\end{tabular}
\centering
\caption{The cell type annotation accuracy across datasets.}
\label{tab:annotation_performance_ours}
\end{table}

We conducted a direct comparison between \algoname~ and several state-of-the-art baseline models. As summarized in Table~\ref{tab:annotation_performance_ours}, \algoname~ outperforms scGPT by 1.0\% on hPancreas, 0.9\% on MS, 1.7\% on Myeloid, and 3.0\% on Myeloid\_b. Compared to GeneMamba, \algoname~ achieves improvements of 1.0\% on hPancreas, 7.4\% on MS, and 1.2\% on Myeloid\_b.
These consistent gains in accuracy demonstrate the effectiveness of \algoname~ in leveraging both global and local contextual information for more accurate and robust cell type annotation. 

\subsection{Multi-omic integration}
The multi-omic integration task \cite{cui2024scgpt} aims to unify data from different molecular modalities (e.g., transcriptome, epigenome, proteome) into a shared embedding space that reflects true biological states rather than technical variation. To evaluate the integration capability of \algoname, we conducted experiments on three representative public datasets \cite{he2024integrated, cui2024scgpt}: 10x Multiome PBMC (RNA + ATAC), BMMC (RNA + protein), and the more challenging mosaic ASAP PBMC dataset.

\begin{table}[ht]
\centering

\begin{tabular}{@{}llc@{}}
\toprule
\textbf{Datasets} & \textbf{Models} & \textbf{Avg\_bio} \\
\midrule
\multirow{5}{*}{10x Multiome PBMC} & scGPT        & 0.758 \\
                                  & scGLUE       & 0.747 \\
                                  & Seurat v4    & 0.722 \\
                                  & scMoMaT      & 0.725 \\
                                  & \textbf{Ours}        & \textbf{0.789} \\
\midrule
\multirow{5}{*}{BMMC (RNA + Protein)} & scGPT        & 0.697 \\
                                  & scGLUE       & 0.600 \\
                                  & Seurat v4    & 0.650 \\
                                  & scMoMaT      & 0.630 \\
                                  & \textbf{Ours}        & \textbf{0.721} \\
\midrule
\multirow{5}{*}{ASAP PBMC}        & \textbf{scGPT}        & \textbf{0.587} \\
                                  & scGLUE       & 0.561 \\
                                  & Seurat v4    & 0.541 \\
                                  & scMoMaT      & 0.546 \\
                                  & Ours         & 0.579 \\
\bottomrule
\end{tabular}
\caption{Comparison of multi-omic integration across datasets.}
\label{tab:multiomic_integration_ours}
\end{table}

We adopted the average biological variance preservation score (Avg\_bio) as the main evaluation metric, where higher values indicate better retention of biological signal post-integration. We compared \algoname~ against leading methods, including scGPT \cite{cui2024scgpt}, scGLUE \cite{cao2022multi}, Seurat v4 \cite{hao2021integrated}, and scMoMaT \cite{zhang2023scmomat}.

As shown in Table~\ref{tab:multiomic_integration_ours}, \algoname~ achieves state-of-the-art performance, with gains of 3.1\% on Multiome PBMC and 2.4\% on BMMC over the best baselines. These results highlight \algoname’s ability to capture both local and global biological context, enabling more accurate and biologically meaningful multi-omic integration.

\section{Conclusion}
In this study, we introduce \algoname, a lightweight hybrid neural network that effectively balances performance and efficiency for pan-cancer single-cell representation learning.Our model introduces a sophisticated hierarchical "local-to-global" design paradigm. This design successfully decouples the complex task of transcriptome modeling into two independent sub-tasks: parallel local context encoding and efficient global information integration.To rigorously evaluate our model, we constructed a large-scale pan-cancer single-cell benchmark, \algoname Bench, spanning 33 cancer subtypes and over 3.5 million high-quality cells. Extensive experiments demonstrate that \algoname~outperforms existing state-of-the-art models across a wide range of tasks, including cell-type annotation, batch correction, gene regulatory network inference, and multi-omic integration. These results highlight the potential of \algoname~as a foundational model for advancing precision oncology and single-cell computational biology.
\section{Acknowledgments}
This work was supported by the National Natural Science Foundation of China (Grant No. 62271237), the Key Project of the Natural Science Foundation of Jiangxi Province (Grant No. 20242BAB26014), the 2025 Pilot Project of Jiangxi Future Health Interdisciplinary Center (Grant No. 9167-28280006), the Hunan Natural Science Foundation (Grant No. 2025JJ50338), and the Shanghai Education Committee AI Project (Grant No. JWAIYB-2).

\bibliography{aaai2026}

\end{document}